# On the relationships between kinetic schemes and two-state single molecule trajectories

Ophir Flomenbom, and Joseph Klafter

*School of Chemistry, Raymond & Beverly Sackler Faculty of Exact Sciences, Tel Aviv University, Ramat Aviv, Tel Aviv 69978, Israel*

**Abstract**

Trajectories of a signal that fluctuates between two states which originate from single molecule activities have become ubiquitous. Common examples are trajectories of ionic flux through individual membrane-channels, and of photon counts collected from diffusion, activity, and conformational changes of biopolymers. By analyzing the trajectory, one wishes to deduce the underlying mechanism, which is usually described by a multi-substate kinetic scheme. In previous works [O. Flomenbom, J. Klafter, and A. Szabo, Biophys. J., in press (2005); O. Flomenbom and J. Klafter, Acta Physica Polonica B **36**, 1527 (2005)], we divided kinetic schemes that generate two-state trajectories into two types: *reducible* schemes and *irreducible* schemes. A full characterization of the reducible ones was given. We showed that all the information in trajectories generated from reducible schemes is contained in the waiting time probability density functions (PDFs) of the two states. It follows that reducible schemes with the same waiting time PDFs are not distinguishable; namely, such schemes lead to identical two-state trajectories in the statistical sense. In this work, we further characterize the topologies of kinetic schemes, now of irreducible schemes, and further study two-state trajectories from the two types of scheme. We suggest various methods for extracting information about the underlying kinetic scheme from the trajectory (e. g., calculate the binned successive waiting times PDF and analyze the ordered waiting times trajectory), and point out the advantages and disadvantages of each. We show that the binned successive waiting times PDF is not only more robust than other functions when analyzing finite trajectories, but contains, in most cases, more information about the underlying kinetic scheme than other functions in the limit of infinitely long trajectories. For some cases however, analyzing the ordered waiting times trajectory may supply unique information about the underlying kinetic scheme.

**I. Introduction**

Since the first patch clamp measurements[1], single molecule experiments have attracted the attention of researchers due to the opportunity they provide in studying complex processes in biology, chemistry and physics in great detail[2-29]. Examples include the flux of ions through individual channels[1-2, 22-25], the translocation of single stranded DNA and RNA through individual nanopores[26-27], diffusion of single molecules[5-9], conformational fluctuations of biopolymers[10-16], single enzyme activity[17-21], and blinking of nano-crystals [28-29]. By observing processes on the single molecule level one wishes to get detailed information about the underlying mechanism, information that cannot be obtained, in most cases, from bulk experimental output. Usually, the underlying mechanism is described by a multi-substate kinetic scheme[1-2, 16-25, 30-33] (for a more involved model describing single molecule activity see, for example, Ref. 34). In many single molecule experiments the observable at the instantaneous time *t* flips between two distinct values, implying that each substate in the underlying kinetic scheme belongs to one of two possible states, which are labeled the *on* and the *off* states. The flipping events produce a two-state trajectory, which is a time-series made of *on* and *off* waiting times (Figs. 1A & 1B). In experiments, due to noise, fluctuations occur around the values of the *on* and the *off* states. The ability to restore reliably the noiseless trajectory from the experimental output depends *roughly* on the sum of the mean fluctuation amplitudes in the observable value in each of the states relative to the difference between the mean of these values. For a recent work that deals with the number of photon counts collected per time interval in single molecule measurements based on the Förster resonance energy transfer mechanism see Ref. 35. Here we are interested in obtaining as much information as possible about the underlying kinetic scheme by analyzing the *noiseless* two-state trajectory generated by the kinetic scheme. In a multi-substate scheme, the number of substates in each of the states can be different (Figs. 2A-2B), the connectivity between substates within a state and between states can be complex, i.e. exceed the one-dimensional nearest neighbors connectivity within a state (Fig. 2C), and can contain a complex network of connections between substates of different states (Figs. 2D-2E). In addition, the scheme may show a net flux in steady state along some connections (i. e. a non-equilibrium steady state), when an external source of energy is present[36].

The central question that arises when trying to go back from the *two-state* trajectory to the *multi-substate* kinetic scheme[17-25, 32-33, 37-48] is: *how much can one learn about the underlying multi-substate kinetic scheme by analyzing two-state single molecule trajectories*? In previous works[32-33] we classified kinetic schemes according to the existence or lack of correlations between successive waiting times in the time-series they generate. Kinetic schemes that lead to uncorrelated waiting times trajectories were termed *reducible*, whereas those that lead to correlated waiting times trajectories were termed *irreducible*. A scheme is reducible due to symmetry, which originates from a special choice of the scheme details, or due to topology (specific combinations of symmetry and topology lead to the same result[33]). By giving the topologies of reducible kinetic schemes, we established a relationship between a general property of the trajectory and the characteristics of the underlying mechanism. An important consequence of our classification is that it is impossible to discriminate between different reducible kinetic schemes that have the same waiting time probability density functions, which are the basic functions that characterize the trajectory. In this paper, we further characterize the topologies of kinetic schemes, now of irreducible schemes, and further study two-state trajectories from both scheme types. We suggest several ways to analyze the time-series. These include calculating the binned successive waiting times PDF, and analyzing the ordered waiting times trajectory. Studying the advantages and disadvantages of each, we show that, in most cases, the binned successive waiting times PDF is not just more robust than other functions when analyzing finite trajectories, but is more informative than other functions in the infinitely long trajectory limit. In some cases however, analyzing the ordered waiting times trajectory may supply unique information about the underlying kinetic scheme.

## II. Reducible and irreducible kinetic schemes

The basic characterization of the time-series is given by the waiting time PDFs of the *on* state, $\phi_{on}(t)$, and of the *off* state, $\phi_{off}(t)$. $\phi_{on}(t)$ and $\phi_{off}(t)$ are obtained from a trajectory by building the histograms from the random *on* state and *off* state waiting times, respectively. These PDFs cannot, in principle, be obtained from bulk measurements. After extracting $\phi_{on}(t)$ and $\phi_{off}(t)$ from the experimental trajectory, one adjusts the details of a multi-substate scheme such that the calculated $\phi_{on}(t)$ and $\phi_{off}(t)$ are the same as the experimental ones. As $\phi_{on}(t)$ and $\phi_{off}(t)$ are the first passage time (FPT) PDFs of the multi-substate scheme decoupled into irreversible *on* and *off* processes with initial conditions being the normalized steady state flux of the coupled system[39], one can always calculate these PDFs given a kinetic scheme. However, when $\phi_{on}(t)$ and $\phi_{off}(t)$ are multi-exponentials, one can find several different schemes that lead to the same $\phi_{on}(t)$ and $\phi_{off}(t)$. This degeneracy raises the question whether one can discriminate between kinetic schemes that lead to the same waiting time PDFs by calculating other functions from the trajectory. These functions include: (*a*) the PDFs of two successive waiting times[16-17, 32-33, 38-43], $\phi_{x,y}(t_1,t_2)$ $x,y = on, off$ ; (*b*) the *x-y* propagator $G(x\,\tilde{t}\,|\,y\,\tau) = G(x\,t\,|\,y\,0)$ [17-19, 21, 41 44-47], which is the bulk relaxation function. Here, $t \equiv \tilde{t} - \tau \geq 0$ and the equality is valid for stationary processes as we consider here; (*c*) higher order propagators[19, 44, 48], e. g. $G(x\,t;\,y\,\tau\,|\,z\,0)$, where $x,y,z = on, off$ and $t \geq \tau \geq 0$; (*d*) PDFs of higher order successive waiting times, e. g. $\phi_{x,y,z}(t_1,t_2,t_3)$. Note that the functions in (*a*), (*c*) and (*d*) can be obtained only from single molecule trajectories.

**II.1 Reducible Schemes**

Reducible schemes are those for which each of the functions (*a*)-(*d*) mentioned above obtained from the trajectory are given in terms of $\phi_{on}(t)$ and $\phi_{off}(t)$.

This means that all the information in the trajectory is already contained in these PDFs. Thus, such a trajectory can be generated from a two-state semi-Markov (TSSM) process (Fig. 2F), with the waiting time PDFs $\phi_{on}(t)$ and $\phi_{off}(t)$. A TSSM process is a process where the *on* and the *off* waiting times are drawn randomly and independently out of non-exponential waiting time PDFs[49-50]. We refer to schemes that generate uncorrelated waiting times trajectories as reducible, because, as far as can be deduced from a trajectory, the complex topology of the scheme is reduced to the simplest topology of a two-state scheme (Fig. 2F). It follows that two-state trajectories from reducible schemes with the same waiting time PDFs are identical in the statistical sense. Namely, it is impossible to distinguish between reducible schemes with the same waiting time PDFs, if all the information about the process is extracted only from a trajectory.

A test for the lack of correlations in the two-state trajectory is the factorization of $\phi_{x,y}(t_1, t_2)$ $x, y = on, off$, into the product of $\phi_x(t_1)$ and $\phi_y(t_2)$ for *every* $x, y = on, off$,

$$\phi_{x,y}(t_1, t_2) = \phi_x(t_1)\phi_y(t_2) \quad ; \quad x, y = on, off. \tag{1}$$

Generally, Eq. (1) holds when the scheme possesses gateway substates in either of the states, or in both. A substate in state *x* is a gateway substate when each event in the state either starts at this substate (type 1) or terminates through it (type 2). If a scheme has one gateway substate (of any type) in either the *on* or the *off* states, then Eq. (1) holds for $x = y$, for both cases of $x = on, off$. One gateway substate, however, is not sufficient for the factorization of $\phi_{x,y}(t_1, t_2)$ for the case $x \neq y$; in particular, one gateway substate of type 1 [type 2] in state *x* is sufficient for the factorization of $\phi_{y,x}(t_1, t_2)$ [$\phi_{x,y}(t_1, t_2)$], but not for the factorization of $\phi_{x,y}(t_1, t_2)$ [$\phi_{y,x}(t_1, t_2)$], see Ref.

33 for the mathematical treatment. A scheme is reducible due to its topology only if it has in addition to a gateway substate[32-33]: (*i*) another gateway substate of a different type in the same state (specific examples are those schemes that have one substate in either the *on* or the *off* states, Figs. 2A-2B and Figs. 3A-3C. A more general example is shown in Fig. 2C); or (*ii*) & (*iii*) another gateway substate of the same type in the other state (Figs. 2D and 2E). We emphasize that since our argument relies on the connectivity of the scheme, cases (*i*)-(*iii*) can be characterized by any waiting time PDF for a substate, and not only by the Markovian (exponential) type[51]. Furthermore, we note that some topologies that correspond to case (*i*) lead to equilibrium at steady state (those schemes that have one substate in either the *on* or the *off* states), whereas those that correspond to cases (*ii*) and (*iii*) lead to non-equilibrium at steady state. To summarize the above possibilities we state that the classes of schemes that fulfill Eq. (1) due to topology are those schemes for which each *on* (*off*) event along the trajectory has the same initial probabilities among the *on* (*off*) substates as the previous *on* (*off*) events. It should be noted, however, that other less general schemes might be reducible when choosing the scheme details in a special way that leads to symmetry[1,2] [see the discussion below Eq. (10)].

To demonstrate the relationship between the scheme topology and the characteristics of the trajectory, as well as the equivalence of trajectories from reducible kinetic schemes, we consider the two schemes shown in Fig. 2A and Fig. 2B (hereafter, schemes 2A and 2B). Both schemes contain *n off* substates and one *on* substate. Specifically, we assume that both processes are characterized by a set of transition rates. The expression for $\phi_{on}(t)$ for scheme 2A reads,

$$\phi_{on}(t) = \Lambda e^{-\Lambda t} \qquad ; \qquad \Lambda = \sum_{j=1}^{n} a_{jon}, \tag{2}$$

where $a_{ion}$ is the transition rate from the *on* substate to the $i^{th}$ *off* substate. The expression for $\phi_{on}(t)$ for scheme 2B is equivalent to that given by Eq. (2), and we can choose the transition rate from the *on* substate to the *off* substate (denoted substate *1*) to be equal to $\Lambda$. The expression for $\phi_{off}(t)$ for scheme 2A is given as a sum of weighted exponentials,

$$\phi_{off}(t) = (1/\Lambda)\sum_{j=1}^{n} a_{jon}a_{onj}e^{-a_{onj}t} \equiv \sum_{j=1}^{n} W_j^{off}\psi_j(t) ; \qquad W_j^{off} = a_{jon}/\Lambda, \qquad (3)$$

where $a_{oni}$ is the transition rate from the $i^{th}$ *off* substate to the *on* substate, and $\psi_i(t) = a_{oni}e^{-a_{oni}t}$ is the waiting time PDF of the $i^{th}$ *off* substate. The expression for $\phi_{off}(t)$ for scheme 2B is also given by a sum of weighted exponentials, and can be made the same as $\phi_{off}(t)$ of scheme 2A. This mapping is accomplished by comparing the Laplace transforms ($\bar{g}(s) = \int_0^\infty g(t)e^{-st}dt$) of $\phi_{off}(t)$ of the two schemes. Specifically, we equate coefficients of similar powers of the Laplace variable *s* in the nominator and the denominator, and then solve the obtained set of *n* equations that relate the transition rates of one scheme to the transition rates of the other scheme. Note that the mapping leads to relationships between the *off* substate transition rates of scheme 2B and the *on* substate transition rates on scheme 2A. Having matched the two waiting time PDFs of the two schemes, we turn now to generating the trajectories. We first generate a trajectory from scheme 2A. A random time is drawn out of $\phi_{on}(t)$ and then a direction is chosen that determines to which *off* substate the process evolves. This stage, however, can be viewed as part of the *off* event (in this Gillespie[52] kind of algorithm, the choice of the direction does not 'cost' time). Thus, the *off* waiting time is generated by first choosing a specific substate *i* according to the weights $\{W_j^{off}\}_{j=1}^n$, and then drawing a random time out of $\psi_i(t)$. This algorithm leads to independence between successive *on-off* waiting times. The next cycle is generated in exactly the same way; namely, the cycle starts from the same single *on* substate. This means that successive *on-on* and *off-off* waiting times are independent, as well. Due to this independence, other algorithms can be used for generating the random waiting times; in particular, each *off* waiting time can be generated using the rejection method[53]. Now, looking at scheme 2B, we notice that due to the scheme special connectivity between the *on* and the *off* substates, each event always starts at the same substate and terminates through the same substate. This leads to independence between each pair of successive waiting times in the trajectory generated by scheme 2B as well, which again means that several algorithms can be used to generate the trajectory. Thus, we can choose the same algorithm to generate trajectories from schemes 2A and 2B. Finally, as we made $\phi_{on}(t)$ and $\phi_{off}(t)$ of the two schemes the same, the trajectories from the two different schemes will have the same statistical properties, and clearly cannot be distinguished.

Technically, to identify a reducible kinetic scheme from a trajectory one should check whether Eq. (1) holds. In practice, $\phi_{x,y}(t_1,t_2)$ is built from the experimental trajectory by constructing a two dimensional histogram of the intersection of successive *x* and *y* waiting times. However, when $\phi_{x,y}(t_1,t_2)$ calculated from the trajectory is too 'noisy' due to the spreading of the limited number of events in the trajectory onto 2 dimensions, another test that discriminates reducible schemes from irreducible ones can be applied. Albeit less informative (see the discussion in section II.2), this test utilizes the *x-y* propagator $G(x\,t\,|\,y\,0)$, $x,y = on, off$. $G(x\,t\,|\,y\,0)$, or the corresponding state-correlation function, can be calculated directly from the time-series, and then compared to the theoretical expression for a TSSM process[54] with the input waiting time PDFs being the experimentally obtained $\phi_{on}(t)$ and $\phi_{off}(t)$ [21]. If the two functions coincide, the scheme that generated the trajectory is reducible. Note that given the basic waiting time PDFs, every function calculated from the trajectory can be compared with the corresponding theoretical function for a renewal process, and thus can be used for discriminating reducible from irreducible kinetic schemes. See section III for an example.

**II.2 Irreducible schemes**

Irreducible schemes are those for which Eq. (1) does not factorize for at least one combination of $x, y = on, off$. We consider two options: (*A*) $\phi_{x,y}(t_1, t_2)$ factorizes for 3 combinations of $x, y = on, off$, and (*B*) $\phi_{x,y}(t_1, t_2)$ factorizes only for $x = y$. The occurrence of case (*A*) can indicate that the kinetic scheme possesses one gateway substate in either state (Fig. 3D), although special symmetric schemes can lead to similar results. The occurrence of case (*B*) can indicate that the kinetic scheme possesses an intermediate gateway substate in either state (Fig. 3E). A substate is an intermediate gateway substate when every event passes through it but does not start or terminate through it. Intermediate gateway substates do not lead to reducible schemes even when they appear in both states or with a gateway substate.

In the remaining of the paper, we study several methods for analyzing trajectories. These methods are mainly useful for extracting information about the details of irreducible schemes, but can be used also to help identify the type of the scheme. By applying the various methods on several trajectories, we characterize the relative advantages and disadvantages of each in supplying as much information as possible about the scheme details. For this we consider the simplest irreducible scheme, which is the four-substate scheme shown in Fig. 3F, hereafter scheme 3F. A two-substate scheme (Fig. 2F), as well as all three-substate schemes are reducible (Figs. 3A-3C). We start by constructing the waiting time PDFs for scheme 3F. The scheme is defined by the waiting time PDF per substate $\psi_i(t)$, $i=1, 2, 3, 4$, and the transition probabilities $\omega_{ji}$ ($\omega_{ji}$ is the transition probability from substate $i$ to $j$). The expression for $\phi_{on}(t)$, which is given in terms of the *on* substate waiting time PDFs $\psi_1(t)$ and $\psi_4(t)$, reads,

$$\phi_{on}(t) = W_1^{on}\psi_1(t) + W_4^{on}\psi_4(t). \tag{4}$$

The weights $W_i^{on}$'s are the probabilities to start an *on* event at substate *i*, and are given in terms of the transition probabilities,

$$W_1^{on} = \omega_{12}\omega_{23} / (\omega_{12}\omega_{23} + \omega_{32}\omega_{43}), \tag{5}$$

and $W_4^{on} = 1 - W_1^{on}$ due to the normalization condition. The expression for $\phi_{off}(t)$ reads,

$$\phi_{off}(t) = W_2^{off} F_2^{off}(t) + W_3^{off} F_3^{off}(t), \tag{6}$$

where due to the connectivity of the scheme we have $W_2^{off} = W_1^{on}$, and $W_3^{off} = 1 - W_2^{off}$. $F_i^{off}(t)$, which is the conditional FPT PDF to exit the *off* state given that the *off* event started at the *off i* substate, is given by,

$$F_i^{off}(t) = f_{1i}^{off}(t) + f_{4i}^{off}(t). \tag{7}$$

$f_{ji}^{off}(t)$ is the conditional FPT PDF to reach substate *j* of the *on* state given that the event started at substate *i* of the *off* state. The Laplace transform of $f_{ji}^{off}(t)$ is calculated by counting all possible trajectories that started at substate *i* of the *off* state and terminated at substate *j* of the *on* state[50], and leads to,

$$\bar{F}_2^{off}(s) = \frac{\bar{\psi}_2(s)\omega_{12}}{\bar{D}(s)} + \frac{\bar{\psi}_2(s)\omega_{32}\bar{\psi}_3(s)\omega_{43}}{\bar{D}(s)}, \qquad (8)$$

and

$$\bar{F}_3^{off}(s) = \frac{\bar{\psi}_3(s)\omega_{23}\bar{\psi}_2(s)\omega_{12}}{\bar{D}(s)} + \frac{\bar{\psi}_3(s)\omega_{43}}{\bar{D}(s)}. \qquad (9)$$

The first (second) term on the expressions for $\bar{F}_i^{off}(s)$ is $\bar{f}_{1i}^{off}(s)$ [$\bar{f}_{4i}^{off}(s)$], where $(\bar{D}(s))^{-1} = (1 - \bar{\psi}_2(s)\omega_{32}\bar{\psi}_3(s)\omega_{23})^{-1}$ is the factor that represents all possible number of transitions between substates 2 and 3 before leaving the *off* state for the first time.

For scheme 3F, the calculations of the two successive waiting times PDF $\phi_{x,y}(t_1, t_2)$, and higher order ones, are straightforward[55]. For example, the difference *off-off* successive waiting times PDF, $\Delta\phi_{off,off}(t_1, t_2) = \phi_{off,off}(t_1, t_2) - \phi_{off}(t_1)\phi_{off}(t_2)$, is given by,

$$\Delta\phi_{off,off}(t_1, t_2) = W_2^{off} W_3^{off} \left(F_3^{off}(t_1) - F_2^{off}(t_1)\right)\left(F_3^{off}(t_2) - F_2^{off}(t_2)\right)$$

$$+ \left(f_{42}^{off}(t_1) - f_{13}^{off}(t_1)\right)\left(W_2^{off} f_{42}^{off}(t_2) - W_3^{off} f_{13}^{off}(t_2)\right), \qquad (10)$$

and is a symmetric function of the time arguments $t_1$ and $t_2$, as $f_{42}^{off}(t) \propto f_{13}^{off}(t)$. Note that for any reducible scheme $\Delta\phi_{off,off}(t_1, t_2)$ (and more generally $\Delta\phi_{x,y}(t_1, t_2)$, $x, y = on, off$) vanishes by definition. Equation (10) vanishes only for a symmetric choice of the scheme details that leads to $F_3^{off}(t) = F_2^{off}(t)$ and $f_{42}^{off}(t) = f_{13}^{off}(t)$. This means that a symmetric scheme is reducible.

The binned, or summed, waiting times PDF, defined by,

$$\phi_{x+y}(t) = \iint_0^\infty \delta(t - t_1 - t_2)\phi_{x,y}(t_1, t_2)dt_1 dt_2, \qquad (11)$$

and its difference,

$$\Delta\phi_{x+y}(t) = \iint_0^\infty \delta(t - t_1 - t_2)\Delta\phi_{x,y}(t_1, t_2)dt_1 dt_2, \qquad (12)$$

with the Laplace transform relation $\bar{\phi}_{x+y}(s) = \bar{\phi}_{x,y}(s,s)$ and $\Delta\bar{\phi}_{x+y}(s) = \Delta\bar{\phi}_{x,y}(s,s)$, plays a significant role in the analysis of finite trajectories. $\phi_{x+y}(t)$ is obtained from a trajectory by constructing the histogram of the random times: $t_{x+y,i} = t_{x,i} + t_{y,i}$, where $t_{x,i}$ and $t_{y,i}$ are successive waiting times, $i=1,..N$ if $x \neq y$ and $i=1,..N-1$ if $x = y$, and $N$ is the number of *on-off* cycles in the trajectory. We wish to compare this function to other single-argument functions. For this we choose the equal time successive waiting times PDF $\phi_{x,y}(t_1,t_2)\big|_{t_1=t_2 \equiv t}$. Other option for a comparison, which will not be considered here, is the *x-y* propagator $G(x\,t\,|\,y\,0)$. $G(x\,t\,|\,y\,0)$ however is built from not a precise number of *on-off* cycles, so it mixes more strongly the details of the *on* and the *off* substates than, for example, $\phi_{x,y}(t_1,t_2)\big|_{t_1=t_2 \equiv t}$. On the other side, $\phi_{x,y}(t_1,t_2)\big|_{t_1=t_2 \equiv t}$ obtained from a finite trajectory is nosier than $\phi_{x+y}(t)$ and $G(x\,t\,|\,y\,0)$, as it is built out of much less events that consist the trajectory than the other two PDFs. We compare below $\phi_{x+y}(t)$ and $\phi_{x,y}(t_1,t_2)\big|_{t_1=t_2 \equiv t}$ for an infinitely long trajectory generated from scheme 3F for the Markovian case. Thus, we have, $\bar{\psi}_i(s) = \lambda_i/(s+\lambda_i)$, $\lambda_i = \sum_j a_{ji}$, and $\omega_{ji} = a_{ji}/\lambda_i$. We take, $\lambda_1 = 1$, $\lambda_2 = k$, $\lambda_3 = 1$, $\lambda_4 = k$, and $\omega_{12} = \omega_{43} = p$. Here, $k$ sets the extent of asymmetry of the scheme (for $k=1$ the system is symmetric and thus, reducible), and $p$ determines the average number of internal transitions between the *off* substates before a transition *off* → *on* occurs (as $p \to 1$ no such internal transitions are expected to occur). Figures 4A-4C[56] show $\phi_{off,off}(t) = \phi_{off,off}(t,t)$, $\phi_{off}^2(t)$, and $\Delta\phi_{off,off}(t) = \Delta\phi_{off,off}(t,t)$, for $k$ = 0.01, 0.05, and $p$ = 0.35. For this value of $p$ several transitions between the *off* substates are expected to occur in each *off* event. $\phi_{off,off}(t)$, $\phi_{off}^2(t)$, and $\Delta\phi_{off,off}(t)$ show similar behavior in the examined range of parameter values. $\Delta\phi_{off,off}(t)$ exhibits a sharp decay from an initial amplitude, which is followed by a peak appearing at larger times, see Refs. 38-39 for a similar behavior. The peak is two orders of magnitude smaller than the maximal value of the PDF (Figs. 4B-4C). The same qualitative behaviors of $\phi_{off,off}(t)$, $\phi_{off}^2(t)$, and $\Delta\phi_{off,off}(t)$ are obtained for $k = 0.01, 0.05$, and $p = 0.95$ (Figs. 5B-5C), where for this value of $p$ transitions between the *off* substates rarely occur in a given *off* event.

In contrast, $\phi_{off+off}(t)$ and $\phi_{off} * \phi_{off} = \iint_0^\infty \delta(t-t_1-t_2)\phi_{off}(t_1)\phi_{off}(t_2)dt_1dt_2$ are more sensitive to changes in the parameter values (Figs. 6A-6C). For *p=0.35* (Figs. 6A-6B), two peaks appear in $\phi_{off+off}(t)$ for both $k$ values, and their amplitudes are comparable. The difference $\Delta\phi_{off+off}(t)$ shows a global maximum followed by a global negative minimum, and its amplitude increases while decreasing $k$. For *p=0.95* (Figs. 6C-6D), as $k$ decreases, the second peak of $\phi_{off+off}(t)$ is separated from the early time peak, shown as a shoulder for $k = 0.05$ and as a small peak for $k = 0.01$, and its amplitude decreases linearly with $k$. For this case, the second peak represents the bunching of slow events in the ordered waiting time trajectory (see Fig. 7A and the discussion in the next section). The difference $\Delta\phi_{off+off}(t)$ shows similar behavior as for the *p=0.35* case, although, here, a second small peak is visible, occurring after the global negative minimum.

From the above analysis it stems that the binned successive waiting times PDF, $\phi_{x+y}(t)$, is not just more accurately obtained from finite trajectories relative to the equal time successive waiting times PDF $\phi_{x,y}(t_1,t_2)\big|_{t_1=t_2=t}$, but that the former PDF is more sensitive to changes in the scheme parameters. The second point can be explained mathematically, when pointing out that $\phi_{x+y}(t)$ contains more information than $\phi_{x,y}(t_1,t_2)\big|_{t_1=t_2\equiv t}$ about the two-dimensional histogram $\phi_{x,y}(t_1,t_2)$, as the former PDF is obtained by integrating over a line of length $t\sqrt{2}$ that intersects the axes of the two-dimensional plane of $\phi_{x,y}(t_1,t_2)$ in the points $(0,t)$ and $(t,0)$, whereas the later PDF is obtained from only the point $(t,t)$ of this plane.

**III The ordered waiting times trajectory**

Another way of presenting the data is to plot vertically the waiting times as a function of their occurrence. The ordered waiting times trajectory may show pronounced patterns that can be used to obtain valuable information about the scheme type and details. For some cases, the analysis of the ordered waiting times trajectory can be advantageous over other methods. We refer to such a case in the last paragraph of this section.

To study the ordered waiting times trajectory, we first make $\phi_{on}(t)$ and $\phi_{off}(t)$ of the irreducible scheme 3F the same as the corresponding PDFs from the reducible four-substate scheme shown in Fig. 3G (hereafter scheme 3G), by using the same steps mentioned below Eq. (3). The mapping is done for the Markovian case, namely, for an exponential waiting time PDF per substate. The mapping leads to the following relationships between the transition rates of the reducible *off* substates ($b_{21}, b_{12}, b_{32}$) and the irreducible ones,

$$b_{21} = \frac{\lambda_2 \lambda_3 (1-\omega_{32}\omega_{23})}{W_2^{off} a_{12} + W_3^{off} a_{43}}, \tag{13}$$

$$b_{32} = W_2^{off} a_{12} + W_3^{off} a_{43}, \tag{14}$$

$$b_{12} = \lambda_2 + \lambda_3 - b_{21} - b_{32}, \tag{15}$$

and between the transition rates of the reducible *on* substates ($b_{34}, b_{43}, b_{23}$) and the irreducible ones,

$$b_{34} = \frac{\lambda_1 \lambda_4}{W_1^{on} \lambda_1 + W_4^{on} \lambda_4}, \tag{16}$$

$$b_{23} = W_1^{on} \lambda_1 + W_4^{on} \lambda_4, \tag{17}$$

$$b_{43} = W_1^{on} \lambda_4 + W_4^{on} \lambda_1 - b_{34}. \tag{18}$$

Using Eqs. (13-18), it can be easily shown that for every choice of the $a_{ji}$s (> 0 and real) the corresponding $b_{ji}$s are all positive and real as well, namely, such a mapping exist always.

We generate trajectories from the two schemes by using Eqs. (13)-(18), and the same relationships between the transition rates of the irreducible scheme applied in the previous section. We further set $k = 0.1$ and $p = 0.95$. The two trajectories are shown in Figs. 1A and 1B, generated from the irreducible and reducible schemes respectively. The corresponding ordered *off* waiting times trajectories are shown in Fig. 7A and Fig. 7B. Patterns in the ordered *off* waiting times trajectory from the irreducible scheme are immediately noticeable (Fig. 7A), and can be hardly detected in the observable (*on-off*) trajectory (Fig. 1A). The *off* ordered waiting times trajectory generated from the reducible scheme (Fig. 7B) shows no specific patterns. Thus, by looking at the ordered waiting times trajectories one can gain insight into the type of the generating kinetic scheme, an insight that is difficult to obtain from the two-state trajectory. A pronounced pattern in the ordered waiting times trajectory is noticed when at least two distinct groups of waiting times with similar lengths per waiting time in a group are detected. Such patterns are referred to as bunching. Although the waiting times in each of the groups are not correlated, the existence of at least two different groups with common characteristics per event within a group gives rise to correlations in such a trajectory. Thus, we use the term bunching only when the normalized correlation function of the ordered waiting times trajectory, $R_{x,y}(i)$ (see the definition below), is not the Kronecker delta $\delta_{i,0}$, namely, when the two-state process is not renewal[49]. The correlation function of the ordered waiting times trajectory is the same function used by Xie and collaborators[17, 44] and calculated by Cao[39]. Denoting the correlation function of the *off* ordered waiting times trajectory by $R_{off,off}(i)$, it is defined by,

$$R_{off,off}(i-1) = \frac{<t_{off,1}t_{off,i}> - <t_{off}>^2}{<t_{off}^2> - <t_{off}>^2},$$

where $<t_{off}^n> = \int_0^\infty t^n \phi_{off}(t)dt$, $<t_{off,1}t_{off,i}> = \iint_0^\infty t_1 t_i \phi_{off,off}(t_1,t_i)dt_1 dt_i$, and

$\phi_{off,off}(t_1,t_i) = \int ... \int_0^\infty \phi_{off,z_2,...,z_{i-1},off}(t_1,t_2,...,t_{i-1},t_i) \prod_{j=2}^{i-1} dt_j$, ($z_j = off$, $j > 2$). Figure 7C shows $R_{off,off}(i)$ calculated from trajectories of 10000 *on-off* cycles, part of which are shown in Figs. 7A-7B. $R_{off,off}(i)$ from the trajectory generated by the irreducible scheme 3F decays exponentially with $i$ (inset), whereas it is a $\delta_{i,0}$ from the trajectory generated by the reducible scheme 3G. For comparison, Fig. 7D shows $\phi_{off+off}(t)$ calculated from both trajectories. The analytical curves are shown as well. $\phi_{off+off}(t)$ calculated from the 10000 cycles trajectory converges to the analytical curves for both cases. Following the note at the end of section II.1, $\phi_{off+off}(t)$ can be calculated from the trajectory and compared with the theoretical result assuming a renewal process. If the two PDFs coincide the scheme that generated the trajectory is reducible.

Before presenting another way of analyzing the ordered waiting time trajectory, we note that excluding $R_{off,off}(0)$ which is normalized to one, $R_{off,off}(i)$ is obtained from moments of PDFs of different successive *off* waiting times [integrated over the times $t_2,...,t_{i-1}$ ($i>2$)]. In addition, higher order PDFs of successive waiting times are less accurately obtained from finite trajectories. We thus argue that it is hard to get information from $R_{off,off}(i)$ about the scheme details.

Another way of analyzing the ordered waiting times trajectory when bunching occurs is to use a threshold that sets apart the fast from the slow events[57]. More generally, when several timescales are noticeable, several thresholds can be used. In the example shown in Fig. 7A, one can calculate, by using a threshold, the average of the fast *off* waiting times, which is related to the transition rates in the scheme by,

$$\bar{t}_{off,fast} \approx 1/\lambda_2, \tag{19}$$

and the average number of successive fast *off* waiting times, which is related to the scheme transition rates by[58],

$$\bar{n}_{off,fast} + 1 \approx 1/\omega_{32}, \tag{20}$$

where the sign $\approx$ is used to indicate that a threshold value method was applied. Similar expressions are valid for the slower *off* waiting times,

$$\bar{t}_{off,slow} \approx 1/\lambda_3, \tag{21}$$

and

$$\bar{n}_{off,slow} - 1 \approx 1/\omega_{23}. \tag{22}$$

For a trajectory of 10000 events part of which is shown in Fig. 7A, we get by using Eqs. (19)-(22) the following *off* transition rate values: $\lambda_2 = 10.57$, $\lambda_3 = 0.99$, $\omega_{12} = \omega_{43} = 0.08$. These numbers are obtained when applying a threshold value of 2.5, and taking into consideration the value of nearest neighbor waiting times when determining the type of a given waiting time.

The threshold method preformed on the ordered waiting times trajectory is applicable when different timescales are easily detected. This method may give more information about the kinetic scheme than $R_{off,off}(i)$, but not more than $\phi_{off+off}(t)$. For cases where bunching is not detected by looking at the ordered waiting times trajectory, $R_{off,off}(i)$ and $\phi_{off+off}(t)$ can be still calculated. However, for some cases, the signal from $R_{off,off}(i)$ may be very poor, although $\phi_{off+off}(t)$ can be accurately obtained. For example, taking $k = 0.1$ and $p = 0.35$, the ordered waiting times trajectory generated from the irreducible scheme 3F (Fig. 8A) exhibits similar pattern as that generated from a reducible scheme 3G (Fig. 8B). For this choice of parameters, the signal in $R_{off,off}(i)$, which is obtained from a 10000 cycles trajectory, is practically zero (Fig. 8C). However, $\phi_{off+off}(t)$ is still accurately obtained (Fig. 8D). Thus, by using the function $\Delta\phi_{off+off}(t)$ one can determine the type of scheme and to extract information about the scheme details. This example further supports the advantageous of $\phi_{off+off}(t)$ over other methods of analysis.

Nevertheless, the advantage of using the threshold method on the ordered waiting times trajectory when bunching does occur is that it can supply unique information about the kinetic scheme when it contains many substates. For example, in the study of the catalytic activity of individual lipase molecules (lipase B from *Candida antarctica*) bunched fast events were detected in the ordered *off* waiting times trajectory[21]. In this case, $\phi_{off}(t)$ followed a stretched exponential, and the enzymatic activity was modeled by a kinetic scheme with a large number of substates (conformations). Using Eqs. (19)-(20), the average reaction rate of the fast conformations, and the average fluctuation rate from fast to slow conformations were obtained from the ordered *off* waiting times trajectory.

 IV Concluding Remarks

In this paper, we have studied two-state single molecule trajectories generated by multi-substate kinetic schemes. We have been interested in obtaining as much information as possible about the kinetic scheme by analyzing the trajectory. Based on our previous work[32, 33], we have used our general division of kinetic schemes into two groups; *reducible* schemes that generate two-state trajectories with no correlations between waiting times, and *irreducible* schemes that generate correlated waiting times trajectories.

Two-state trajectories from reducible schemes are identical in the statistical sense to trajectories generated by a two-state semi-Markov process with the same waiting time PDFs of the *on* and the *off* states, and are fully characterized by these PDFs. Thus, reducible schemes with the same waiting time PDFs cannot be discriminated by the analysis of a trajectory. The lack of correlations between events along the trajectory stems from special topologies of the underlying kinetic scheme, or indicates for symmetry in the scheme, which results from a special choice of the scheme details (specific combinations of symmetry and topology lead to the same result). To list the special topologies, we have defined a special substate called a gateway substate, where a gateway substate is one in which all events in a given state either start at (type 1) or terminate through (type 2). The topologies that lead to reducible schemes include: (*i*) two gateway substates of different types in either the *on* or the *off* states, and (*ii*) & (*iii*) two gateway substates of the same type in different states.

Two-state trajectories from reducible kinetic schemes supply direct information *only* on the explicit form of $\phi_{on}(t)$ and $\phi_{off}(t)$. From this, one can deduce (to some extent) the number of substates in each of the states, and the scheme connectivity *between* states. Two-state trajectories from irreducible kinetic schemes contain information about the scheme connectivity *within* the states. By calculating the two successive waiting times PDF $\phi_{x,y}(t_1,t_2)$ from the trajectory, which is obtained by constructing a two dimensional histogram of the intersection of successive *x* and *y* waiting times, one can identify an intermediate gateway substate (when Eq. (1) holds only for $x = y$) and one gateway substate (when Eq. (1) holds for exactly three combinations of $x, y = on, off$). An intermediate gateway substate is one where in every event the process passes through but does not start at or terminate through. We note that special symmetric schemes can lead to the same result.

When $\phi_{x,y}(t_1,t_2)$ obtained from the trajectory is too 'noisy' due to the limited number of events in the trajectory, one can construct the binned, or summed, successive waiting times PDF, e. g. $\phi_{x+y}(t)$, $x, y = on, off$, obtained from the trajectory by building the histogram of the random times that are the sum of successive waiting times. $\phi_{x+y}(t)$ has a single-variable, so it is less noisy than $\phi_{x,y}(t_1,t_2)$. $\phi_{x+y}(t)$ contains more information about the scheme than $\phi_{x,y}(t_1,t_2)\big|_{t_1=t_2\equiv t}$, both for technical reasons (only the former PDF is obtained from all successive *x-y* events in the trajectory), and mathematical ones, which makes this PDF more sensitive to changes in the scheme parameters. $\phi_{x+y}(t)$ can be viewed as a more sensitive probe for the scheme details than $G(x\,t\,|\,y\,0)$, because $G(x\,t\,|\,y\,0)$, in contrast to $\phi_{x+y}(t)$, contains information from not a precise number of *x-y* events, so it mixes more strongly the details of the *on* and the *off* substates. These two PDFs, however, can be used for identifying the scheme type: for reducible schemes $\Delta\phi_{x+y}(t) = 0$, and $G(x\,t\,|\,y\,0)$, or the corresponding state-correlation function, coincides with the theoretical one for a TSSM process.

Another way of extracting information from the time-series is obtained by analyzing the ordered waiting time trajectory. This is the trajectory of the waiting times plotted vertically, either only *on* or *off* waiting times or *on-off* waiting times, as a function of their chronological occurrence. The ordered waiting times trajectory, which is easily obtained from the data, may display bunching from a relatively small number of events. Bunching means that at least two distinct groups of waiting times with similar length per waiting time in a group are detected in the ordered waiting times trajectory. One can calculate the correlation function of this trajectory, $R_{x,y}(i)$, or to use a threshold for a strong bunching situation to get information about the scheme type and details. We have found that $\phi_{x+y}(t)$ is again a better tool in analyzing the data than $R_{x,y}(i)$ both for technical reasons (for the no visible bunching case, the signal in $R_{x,y}(i)$ is poor, although $\phi_{x+y}(t)$ is still accurately obtained) and mathematical ones ($R_{x,y}(i)$ is the obtained from moments of different order of successive waiting times PDFs). Clearly, $\phi_{x+y}(t)$ is advantageous over the threshold method in the weak bunching limit. However, when the scheme contains many substates, and for a strong bunching situation, the threshold method applied on the ordered waiting time trajectory may supply information that cannot be obtained from other methods.

As a final remark we refer to a case where a trajectory contains more than two detectable states but less than the number of substates in the underlying kinetic scheme. Indeed, such a trajectory will provide more details about the process than a two-state trajectory. However, as it will not represent all the substates of the system, similar ideas to those presented here and in our previous works[32, 33] should be considered when analyzing it.

## Acknowledgments

We thank A. Szabo, A. M. Berezhkovskii, I. Gopich, J. Cao, and J. B. Witkoskie for fruitful discussions. J. K. acknowledges the support of the Israel Science Foundation.

$$\overline{C}(s) = \frac{1}{s}\left[1 - \frac{N}{s}\frac{[1-\overline{\phi}_+(s)][1-\overline{\phi}_-(s)]}{1-\overline{\phi}_-(s)\overline{\phi}_+(s)}\right] \; ; \; N = (\overline{t}_+ + \overline{t}_-)/\overline{t}_+\overline{t}_-, \; \overline{t}_\pm = \int_0^\infty \phi_\pm(t)t\,dt.$$

[55] The expression for the $n \geq 1$ ordered successive *off* waiting times PDF for scheme 3F reads:

$\phi_{off,...,off}(t_n,...,t_1) = W_2^{off} A_n(t_n,...,t_1) + W_3^{off} B_n(t_n,...,t_1)$, where for convenience we denote $t_i \to t_{n+1-i}$. The recursion relations for the $A_n$'s and $B_n$'s read,

$A_n(t_n,...,t_1) = f_{12}^{off}(t_n) A_{n-1}(t_{n-1},...,t_1) + f_{42}^{off}(t_n) B_{n-1}(t_{n-1},...,t_1)$,

$B_n(t_n,...,t_1) = f_{13}^{off}(t_n) A_{n-1}(t_{n-1},...,t_1) + f_{43}^{off}(t_n) B_{n-1}(t_{n-1},...,t_1)$, with $A_0 = B_0 = 1$.

[56] All the analytical curves shown in Figs. 4-8 were obtained using Eqs. (4)-(12) and the expressions in Ref. 55 for the Markovian case, the Laplace transform relation given below Eq. (12), and the standard inverse Laplace transform method.

[57] A more accurate way of determining the type of a waiting time than the threshold method, for which a slow or a fast waiting time is determined by its value relative to the threshold, takes into account the values of two (or more) neighboring waiting times. The simplest algorithm for this transforms the sequence {>, <, >} into the sequence {>, >, >}, and the sequence {<, >, <} into the sequence {<, <, <}. Here > (<) means that the value of the waiting time is larger (smaller) than the value of the threshold.

[58] Calculating $\overline{n}_{off,fast}$ in the limit of strong bunching is equivalent to calculating the average number of successive heads when tossing a biased coin with probabilities $l$ for head and $r$ for tail. The result is $1/r$, which translates to $1/\omega_{32}$ in our case. To get the correct result for $\overline{n}_{off,fast}$ as obtained from the ordered waiting time trajectory one should add a unity to $\overline{n}_{off,fast}$, because in each *fast→slow* transition the first fast

(short) event is masked by a slow (long) one. For the same reason one should subtracted a unity from $\bar{n}_{off,slow}$.

**Figure Captions:**

**Figure 1** Two trajectories of an observable that fluctuates between two values (*on* and *off*) as a function of time. The trajectories are obtained by simulating the kinetic schemes shown in Fig. 3F (**A**) and Fig. 3G (**B**), when making $\phi_{on}(t)$ and $\phi_{off}(t)$ of the two schemes the same (see section III for details).

**Figure 2 A-E** A set of reducible kinetic schemes, and a TSSM scheme (**F**), characterize only by the waiting time PDFs $\phi_{on}(t)$ and $\phi_{off}(t)$. **A** An *n* uncoupled *off* substates connected to one *on* substate. A full arrow between two substates represents a connection in the direction of the arrow. The dashed line represents the *off* substates that are not shown. **B** An *n* coupled *off* substates with one o*n* substate scheme. **C** A reducible scheme with two gateway substates in the same state (the *on* state). The bolded pentagons with full lines stand for a region with any complex network of connections within a state. The dashed arrow stands for a set of connections between many *off* substates and one *on* substate, and the dashed–dotted arrow stands for a set of connections between one *on* substate and many *off* substates. **D-F** When the gateway substates in both the *on* and the *off* states are of type 1 (**D**), or of type 2 (**E**), the scheme is reducible to a TSSM scheme (**F**).

**Figure 3** A set of reducible and irreducible kinetic schemes. For the Markovian case, an arrow from substate *i t*o *j* represents a transition with a rate $a_{ji}$. More generally, an arrow from substate *i* to *j* represents a transition with probability $\omega_{ji}$, where substate *i* has a waiting time PDF $\psi_i(t)$. **A-C** All possible three-substate schemes are reducible. **D** An example for an irreducible scheme with a single gateway substate, denoted *on 3*. **E** An example for an irreducible scheme with a single intermediate gateway substate, denoted *on 3*. **F** The simplest irreducible scheme. **G** A reducible four-substate scheme.

**Figure 4 A** $\phi_{off,off}(t)$ (upper curve), $\phi_{off}^2(t)$ (lower curve), for the four-substate irreducible scheme (Fig. 3F), for *k=0.01, 0.05,* and *p=0.35*. **B-C** The difference $\Delta\phi_{off,off}(t)$. Here, and in all the other figures, the function log(·) stands is the natural logarithm of ·, i.e. ln(·).

**Figure 5 A-C** $\phi_{off,off}(t)$, $\phi_{off}^2(t)$ and $\Delta\phi_{off,off}(t)$ are shown for *k=0.01, 0.05,* and *p=0.95*.

**Figure 6** $\phi_{off+off}(t)$ (top curves) and $\phi_{off} * \phi_{off}$ (**A, C**), and $\Delta\phi_{off+off}(t)$ (**B, D**) are shown for the same range of parameters as in Figs. 4-5. For the top panels (**A-B**) *p=0.35*, and for the bottom panels (**C-D**), *p=0.95*.

**Figure 7 A - B** – The ordered *off* waiting times trajectories that corresponds to the *on-off* trajectories shown in Figs. 1A-1B. The *on-off* trajectories are obtained by simulating the kinetic schemes shown in Fig. 3F (**A**) and Fig. 3G (**B**), when making $\phi_{on}(t)$ and $\phi_{off}(t)$ of the two schemes the same, for scheme 3F (arbitrary units) parameters, $\lambda_1 = \lambda_3 = 1$, $\lambda_2 = \lambda_4 = 0.1$, and $\omega_{12} = \omega_{43} = 0.95$ ($a_{ji} = \lambda_i \omega_{ji}$). **C** – The correlation functions of the *off* ordered waiting times trajectories, calculated from 10000 events trajectories. Excluding the first point which is normalized to one, $R_{off,off}(i)$ decays exponentially for the trajectory from scheme 3F (circle symbol), $R_{off,off}(i) = 0.27e^{-0.12i}$. See the inset plotted on a log-linear scale (also shown is a fitting function). For the trajectory from the reducible scheme 3G (triangle symbols), $R_{off,off}(i) = \delta_{i,0}$. **D** − $\phi_{off+off}(t)$ s calculated from the 10000 events trajectories (circle and triangle symbols as in **C**), and the analytical (dashed) curves. The calculated $\phi_{off+off}(t)$ s converge to the analytical curves. Differences between the $\phi_{off+off}(t)$ s from the two trajectories are detectable.

**Figure 8 A-B** - The ordered waiting time trajectories as in Fig. 7, but for other set of parameters for scheme 3F, $\lambda_1 = \lambda_3 = 1$, $\lambda_2 = \lambda_4 = 0.1$, and $\omega_{12} = \omega_{43} = 0.35$. **C** − The signal in $R_{off,off}(i)$ calculated from a 10000 events trajectory is poor, and the curves from the irreducible scheme 3F and the reducible scheme 3G are practically the same; Namely, for both cases $R_{off,off}(i) \approx \delta_{i,0}$ (see $R_{off,off}(i)$ from the irreducible scheme 3F in the inset, shown on a log-linear scale). **D** − However, $\phi_{off+off}(t)$ s calculated from the 10000 events trajectories converge to the analytical (dashed) curves. For both **C** and **D**, circle and triangle symbols are for the irreducible (3F) and the reducible (3G) schemes.

FIG 1

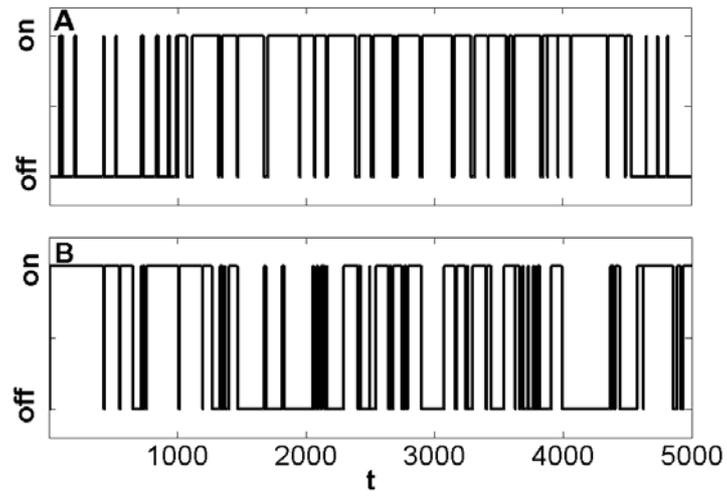

FIG 2

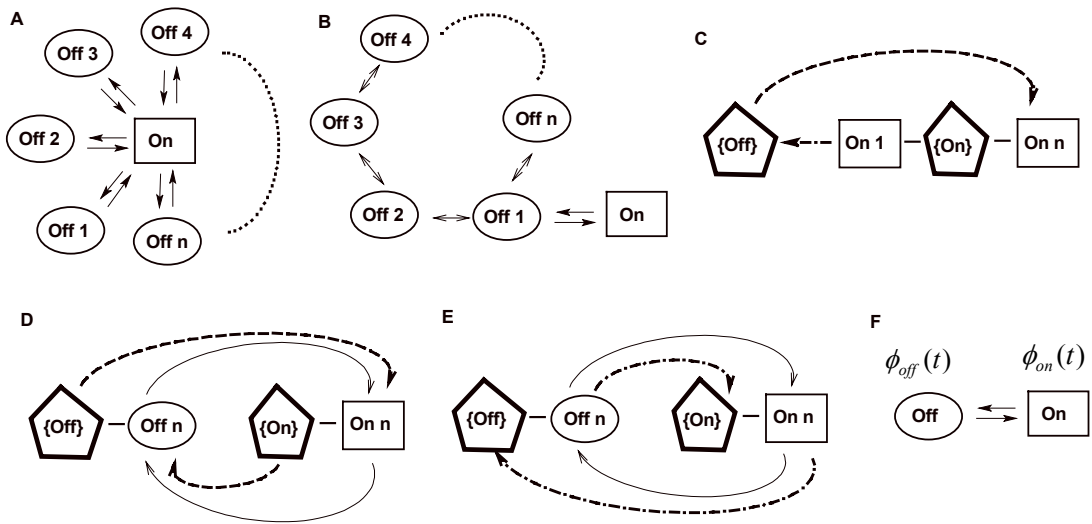

FIG 3

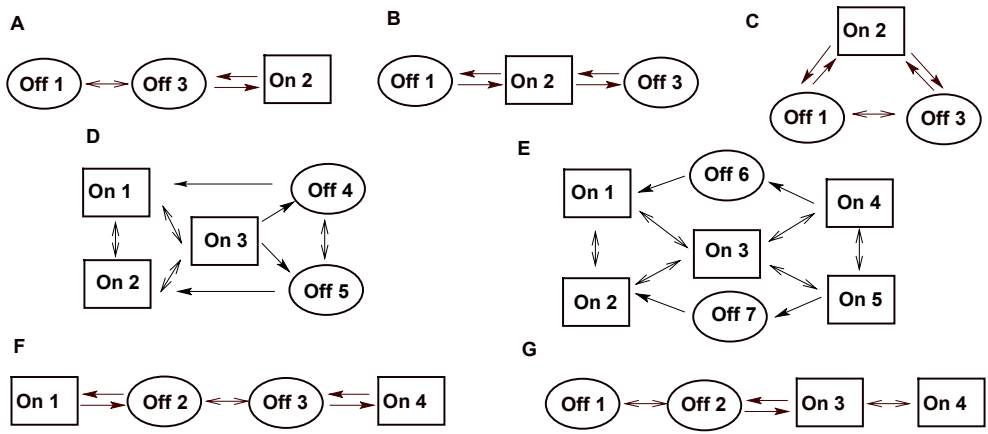

FIG 4

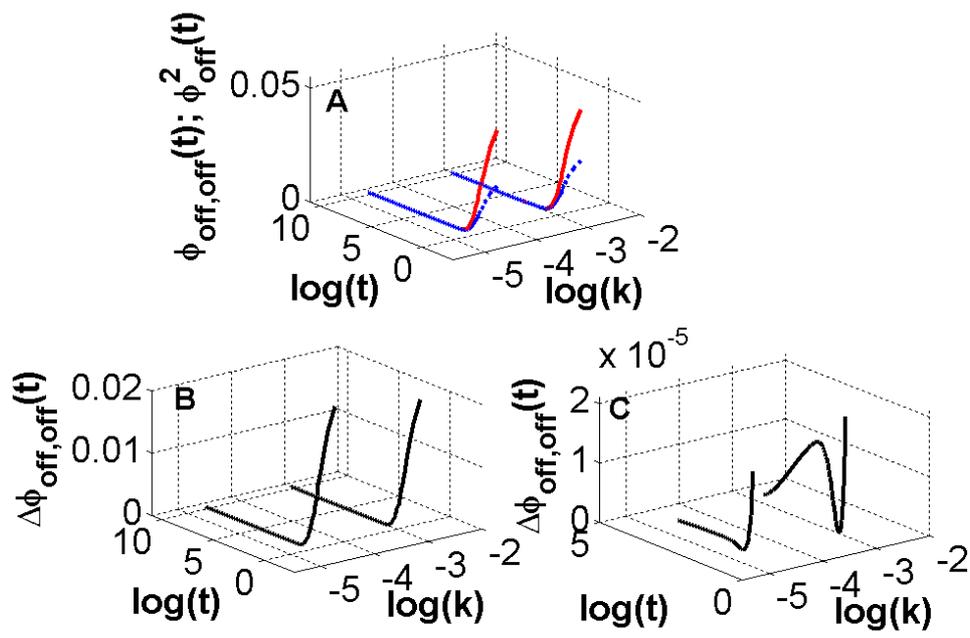

FIG 5

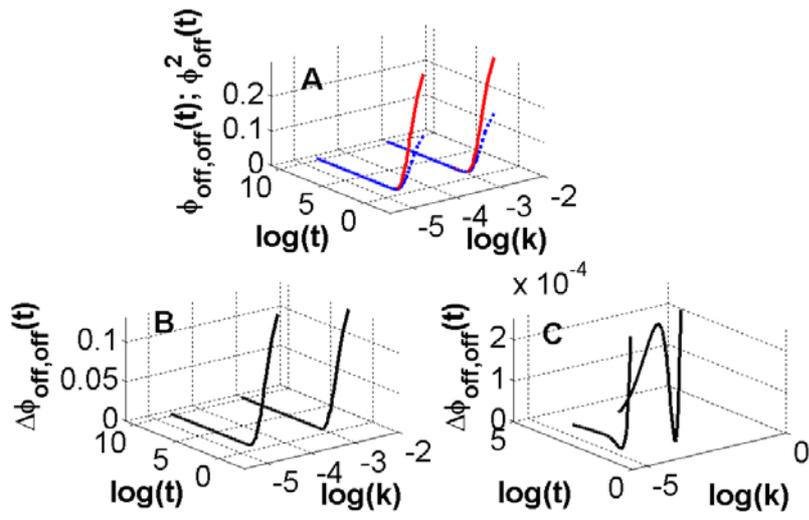

FIG 6

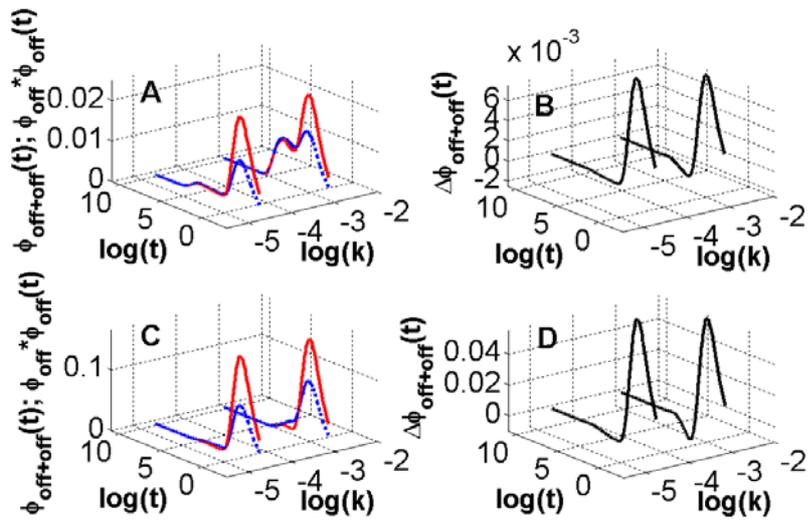

FIG 7

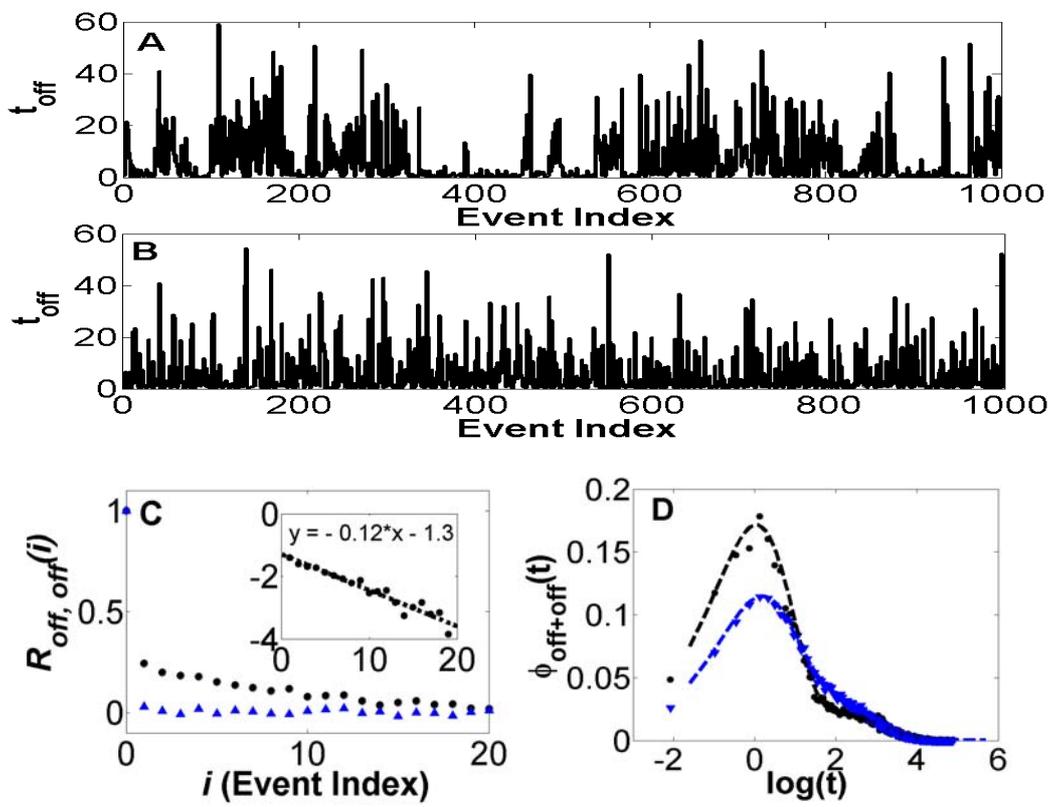

FIG 8

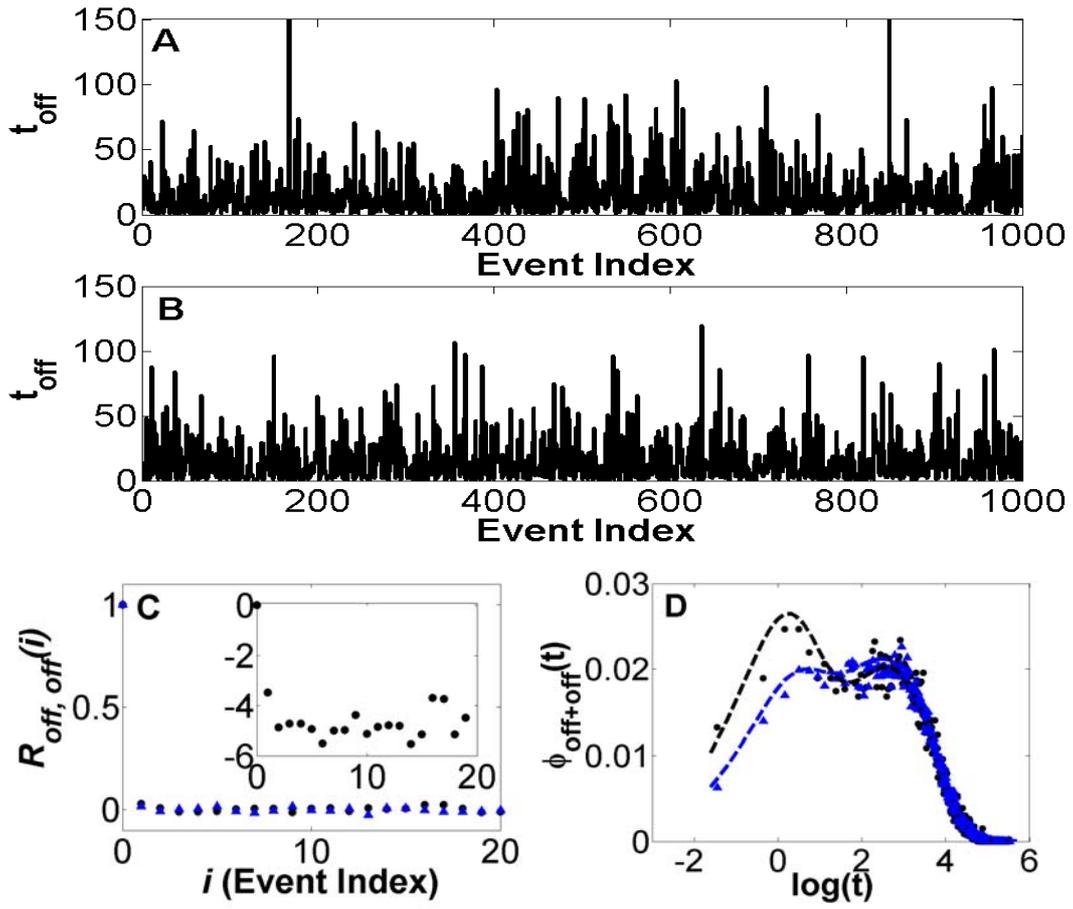